\documentclass[]{aa}
\usepackage{graphicx}
\usepackage{natbib}

\begin{document}
   \title{VLTI/AMBER interferometric observations of the recurrent Nova RS Oph 5.5 days after outburst\thanks{Based on observations made with the Very Large Telescope Interferometer at
       Paranal Observatory under programs 276.D-5049}}

\titlerunning{AMBER observations of the outburst of RS\,Oph}

\authorrunning{Chesneau, O. et al.} 

   \author{O.~Chesneau\inst{1}, N.~Nardetto\inst{1,2}, F.~Millour\inst{3,4}, Ch.~Hummel\inst{5}, A. Domiciano de Souza\inst{3}, D.~Bonneau\inst{1}, M.~Vannier\inst{5}, F.T.~Rantakyr\"o\inst{5},  A.~Spang\inst{1}, F.~Malbet\inst{4}, D.~Mourard\inst{1}, M.F.~Bode\inst{6}, T.J.~O'Brien\inst{7}, G. Skinner\inst{8}, R.~Petrov\inst{3}, Ph.~Stee\inst{1},  E. Tatulli\inst{9}, F.~Vakili\inst{3}}

   \offprints{O. Chesneau}

\institute{Observatoire de la C\^{o}te d'Azur, Dpt. Gemini-CNRS-UMR 6203, Avenue Copernic, F-06130 Grasse, France\\
\email{Olivier.Chesneau@obs-azur.fr}
\and
Max-Planck-Institut f\"ur Radioastronomie, Auf dem H\"ugel
69, 53121 Bonn, Germany
\and
Laboratoire Universitaire d''Astrophysique de Nice, UMR 6525
Univ. de Nice/CNRS, Parc Valrose, F-06108 Nice, France
\and
Laboratoire d''Astrophysique de Grenoble, UMR 5571 Univ.
J. Fourier/CNRS, BP 53, F-38041 Grenoble, France
\and
European Southern Observatory, Casilla 19001, Santiago 19, Chile
\and
Astrophysics Research Institute, Liverpool John Moores University, Birkenhead, CH41
1LD, UK
\and
Jodrell Bank Observatory, School of Physics and Astronomy, Univ. of Manchester,
Macclesfield, SK11 9DL, UK
\and
CESR, 31028 Toulouse, France and Univ. P. Sabatier, 31062 Toulouse, France
\and
Osservatorio Astrofisico di Arcetri, Largo E. Fermi, 5. 50125 Firenze, Italia}
   \date{Received; accepted }

  \abstract
 {}
   {We report on spectrally dispersed interferometric AMBER/VLTI observations of the recurrent nova RS\,Oph five days after the discovery of its outburst on 2006 Feb 12.}
   {Using three baselines ranging from 44 to 86m, and a spectral resolution of $\lambda$/$\delta \lambda$=1500, we measured the extension of the milliarcsecond-scale emission in the K band continuum and in the Br$\gamma$ and He~I~2.06$\mu$m lines, allowing us to get an insight into the kinematics of the line forming regions. The continuum visibilities were interpreted by fitting simple geometric models consisting of uniform and Gaussian ellipses, ring and binary models. The visibilities and differential phases in the Br$\gamma$ line were interpreted using skewed ring models aiming to perform a limited parametric reconstruction of the extension and kinematics of the line forming region.}
   {The limited $uv$ coverage does not allow discrimination between filled models (uniform or Gaussian ellipses) and rings. Binary models are discarded because the measured closure phase in the continuum is close to zero. The visibilities in the lines are at a low level compared to their nearby continuum, consistent with a more extended line forming region for He~I~2.06$\mu$m than Br$\gamma$. The ellipse models for the continuum and for the lines are highly flattened (b/a$\sim$0.6) and share the same position angle (PA$\sim$140$^\circ$). Their typical Gaussian extensions are 3.1x1.9\,mas, 4.9x2.9\,mas and 6.3x3.6\,mas for the continuum, Br$\gamma$ and He~I~2.06$\mu$m lines, respectively. Two radial velocity fields are apparent in the Br$\gamma$ line: a 'slow' expanding ring-like structure ($v_{rad} \leq $1800km.s$^{-1}$), and a 'fast' structure extended in the E-W direction ($v_{rad}\sim$2500-3000km.s$^{-1}$), a direction that coincides with the jet-like structure seen in the radio. These results confirm the basic fireball model, contrary to the conclusions of other interferometric observations conducted by Monnier et al. (2006).}
   {}


   \maketitle
%

\section{Introduction}
RS\,Ophiuchi is a Symbiotic Recurrent Nova, a binary system that comprises a red giant star (RG) and a white dwarf (WD) of mass near the Chandrasekhar limit (see Hachizu \& Kato 2000, 2001, Shore et al. 1996, Bode 1987 and references herein). RS Oph is famous due to its brightness during outburst (m$_V$$\sim$4) and a short recurrence period of about 20yrs. Historically, RS Oph
underwent five outbursts, in 1898, 1933, 1958, 1967, and
1985, with the light curves very similar to each other (e.g.,
Rosino 1987). Recurrent novae are characterized by nova eruptions
with a recurrence time scale from a decade to a century and a brightness amplitude larger than 6 magnitudes,
They show some characteristic differences from classical novae (Starrfield et al. 2000, Gehrz et al. 1998, Gehrz 1988): in particular very short recurrence periods ($\sim$ 10$^2 years$ compared to about 10$^5$ years for classical novae) theoretically require very massive white dwarfs close to the Chandrasekhar mass limit (Hachisu \& Kato 2001). If the mass of the white dwarfs in recurrent novae is assumed to increase, they should eventually explose as a Type Ia supernova when they reach the Chandrasekhar mass limit of about 1.38 solar mass but this hypothesis is still debated (Starrfield et al. 2004, Wood-Vasey \& Sokoloski 2006).

RS\,Oph was discovered in outburst in 2006 Feb. 12.829 UT (defined as outburst time t$_0$, Hirosawa 2006), having undergone its last major outburst in 1985 (see Bode 1987). A campaign was organized that incorporated observations ranging from radio to X-ray wavelengths and an impressive amount of data is being gathered, as is apparent from the recently published papers (Hachizu \& Kato 2006, O'Brien et al. 2006, Sokoloski et al. 2006, Bode et al. 2006, Evans et al. 2006, Monnier et al. 2006a, Hachizu et al. 2006, Das et al. 2006a). In 1985, the radio interferometry technique had reached the sensitivity and spatial resolution necessary to get a picture of the ejection during the earliest stages (Padin et al. 1985, Taylor et al. 1989, Lloyd et al. 1993), and nowadays optical interferometry is mature enough to provide complementary observations of this complex phenomenon.

Some transient phenomena such as novae in outburst have already been studied using optical interferometers e.g. Nova Cyg 1992 (Mark III, Quirrenbach et al. 1993) or recently V838 Mon (PTI, Lane et al. 2005a) and Nova Aql 2005 (V1663 Aql, PTI, Lane 2005b). The Very Large Telescope Interferometer (VLTI) offers the opportunity to investigate the earliest stages of novae from the near-IR with the AMBER instrument (Petrov et al. 2003) and the mid-IR with the MIDI instrument (Leinert et al. 2003). The current outburst of RS Ophiuchi has been followed by four optical interferometers worldwide: IOTA, PTI, KI from t=4 days to t=65 days (Monnier et al. 2006). We note that these optical interferometers lacked spectral dispersion to disentangle the continuum versus emissive line signals. We present AMBER observations of RS Oph secured rapidly, only 5.5 days after the outburst, in the K band with medium spectral resolution (R=1500). The AMBER data provide information on the global shape and for the first time, the kinematics of the ejection in two lines, Br~$\gamma$~2.17$\mu$m and He~I~2.06$\mu$m. The later AMBER observations, performed in H and K bands at low spectral resolution are not presented here. 

All the observations, from the X-rays to the radio domain suggest a complex ejection process, far from the simple view of a spherical ejection. We stress that the set of interferometric observations presented here is limited and it is not  the scope of the paper to constrain the distance of the object. We adopt hereafter the distance of 1.6kpc established by the observations of the last outburst (Hjellming et al. 1986, Snijders et al. 1987, Taylor et al. 1989) and confirmed by the present ones (Bode et al. 2006, Sokoloski et al. 2006,  Hachisu et al. 2006b).

The paper is structured as follows. In Sect.2, we describe the AMBER observations and the data processing. In Sect.3, we characterize the most striking features of the signal by performing a qualitative interpretation. The Sect.4 is devoted to the quantitative interpretation of the absolute visibilities from the continuum only, with simple models limited to three free parameters. This section also includes a discussion of the data obtained from quasi-simultaneous observations of the IOTA and Keck interferometers (Monnier et al. 2006a). In Sect.5, an attempt is made to exploit the rich and complex information of the spectrally dispersed Br$\gamma$ line by means of a family of parametric models sharing spectrally independent properties: a continuum defined in Sect.4 and narrow elliptical 'skewed' rings for the line emission. The aim is to characterize further the kinematics of the line, but this 'parametric imaging' approach is by no means a coherent model of the geometry and kinematics of such a complex source. Lastly, in Sect.6 and Sect.7, we conclude and discuss these optical interferometry observations in the frame of the expanding fireball model, relating them to the large $uv$-coverage radio-interferometry image secured at day 13.8. 

\begin{figure}
  \begin{center}
\includegraphics[width=7cm]{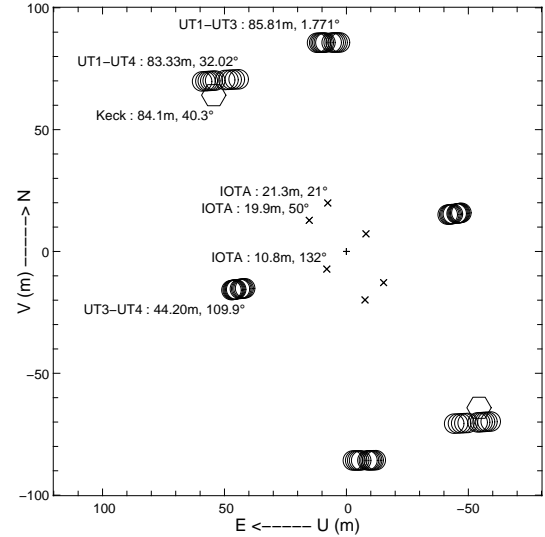}
 \end{center}
 \caption[]{ Mean $uv$ coverage over the time of observations of the VLT baselines. The IOTA and Keck baselines are also shown (Monnier et al. 2006). The projected baseline of the Keck measurement is very close to the U14 baseline. Note that the VLT and Keck observation were performed in K band whereas IOTA ones were performed in H band, leading to an increase of about 30\% in spatial resolution for a given baseline expressed in meter.
\label{fig:uv}}
\end{figure}

\section{Observations}
The VLT telescopes UT1, UT3 and UT4 were used providing mean projected baselines 
of 86.1m (PA=4.5$^\circ$, UT1-UT3, hereafter U13), 46.8m (PA=109.0$^\circ$, U34) and 87.0m (PA=36.0$^\circ$, U14) over the time of observation (see $uv$-coverage in Fig.1). The observations were conducted on 2006 Feb 18. (UT=7h, JD=2,453,784.38), i.e. at t=5.5 days after the outburst. This corresponds to orbital cycle E=7.97 of Fekel et al. (2000) ephemeris: the red giant is at maximum radial velocity and is slightly in front. Only one triplet of visibility measurements was recorded due to the unfavorable position of the nova in the morning sky. Two data sets were recorded: a high SNR record (10000 frames) centered in the Br~$\gamma$~2.166$\mu$m line and the close-by continuum (velocity range -8000/4000\,km.s$^{-1}$) and a low SNR record (30 frames) covering a larger spectral band from 1.95 to 2.28$\mu$m that includes two strong lines, the Br$\gamma$ as well as He~I~2.058$\mu$m. The detector integration time was 50ms as the seeing was good and the spectral resolution was 1500 ($\Delta v=200$km.s$^{-1}$). 

\begin{figure*}
  \begin{center}
\includegraphics[width=17cm]{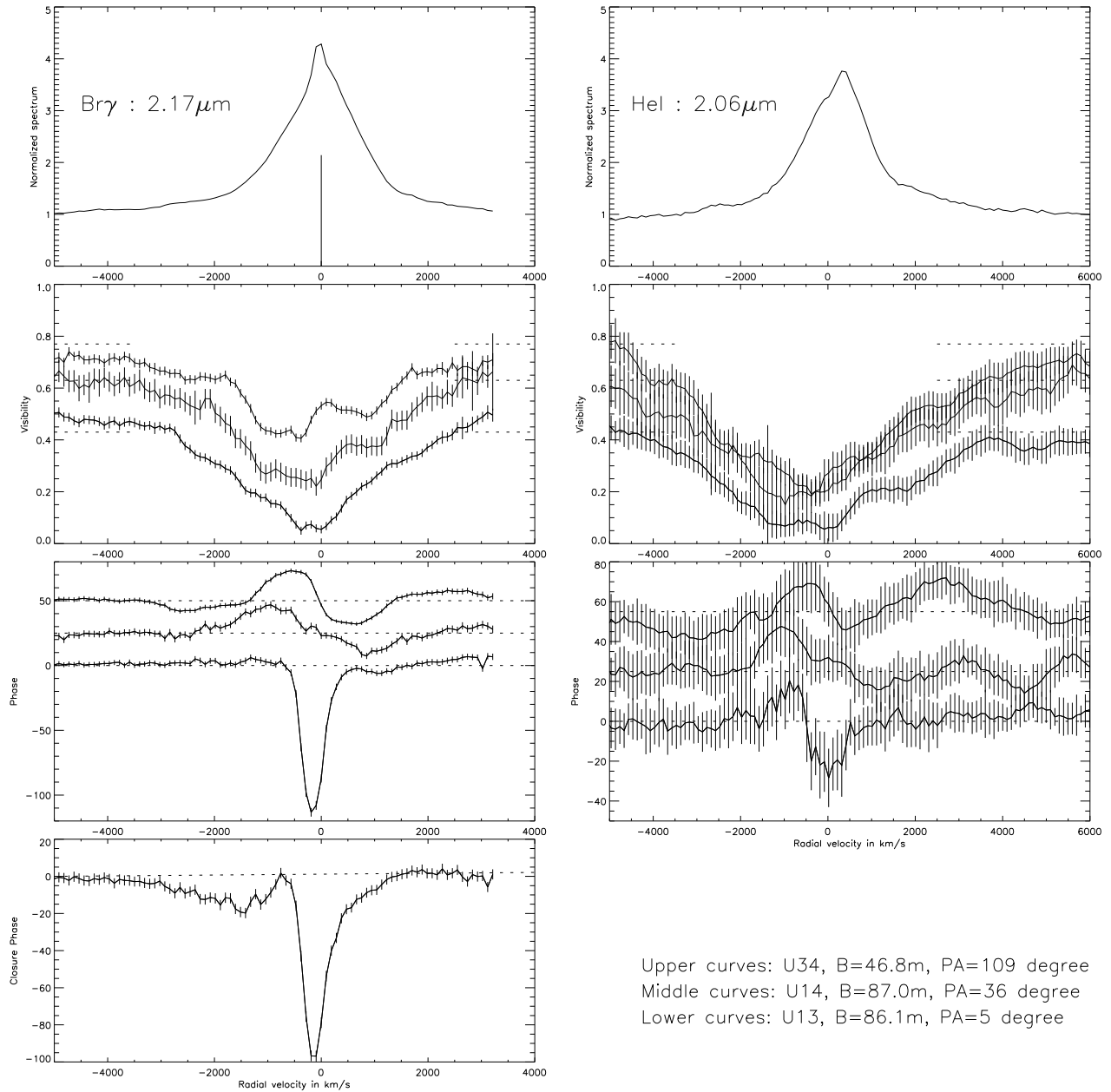}
 \end{center}
 \caption[]{AMBER observables extracted from the data recorded the 2006 February 18: {\bf From top to bottom}: {\bf 1)} AMBER spectrum of the Br$\gamma$ (left) and HeI (right) lines. {\bf 2)} Visibilities for the three baseline (upper line U34, middle U14, lower U13). The mean continuum levels of the high SNR data (dotted lines) are overplotted for comparison. 
{\bf 3)} Differential phases (upper line U34, middle U14, lower U13). The U14 and U13 lines are shifted by 25 and 50 degrees, respectively. The mean continuum levels (dotted lines, zero level) are overplotted for comparison. {\bf 4)} The closure phase. The dotted line emphases the zero level. All the curves result from the interpolation of the measurements by straight lines.
\label{fig:obs}}
\end{figure*}

For the data reduction, we used the software based on the P2VM
(pixel-to-visibilities matrix) algorithm (Tatulli et al. 2006). Three spectrally dispersed visibility and differential phase curves and one spectrum and closure phase curve are derived. The visibilities were corrected for atmospheric and instrumental effects using a calibrating star, $\epsilon$\,Oph (G9.5III, m$_K$=1.1, $\theta$=2.92$\pm$0.03\,mas in CHARM2 catalog Richichi et al. 2005, $\theta$=2.94$\pm$0.08\,mas from Richichi\& Percheron 2005).
For the science source, the error budget includes for each observable the statistical
noise of the observations, which is much lower for the high SNR data compared to the low SNR ones, and for the visibilities the bias from the estimation of the calibrator diameter. The data for Br$\gamma$ (high SNR record) and HeI (low SNR record) are shown in Fig.~\ref{fig:obs}.

 The low SNR curves of Br$\gamma$ (not shown) agree fairly well with the high SNR ones, providing confidence in the HeI results. In order to improve their quality, the low SNR data were smoothed with a 9 pixels box, degrading the spectral resolution to about R=500. The error bars in the closure phase of low SNR data prevented the extraction of any scientifically valid constraint on the object symmetry. The Br$\gamma$ line is dominantly triangular and peaks at about 4.3 times the continuum level. HeI is also triangular but its blue side is more extended. The FWHMs (deconvolved from the spectral resolution of AMBER) of Br$\gamma$ and HeI are about 1300\,km.s$^{-1}$ and 1500\,km.s$^{-1}$, respectively (see also Evans et al. 2006, Das et al. 2006a).

\section{Qualitative interpretation}

In the results shown in Fig.2, it is immediately obvious that the emission from the nova ejecta and environment are not spherical.
The continuum is well resolved and the baseline U13 (V$_{cont}$=0.5$\pm$0.02) shows a visibility in the continuum significantly smaller than baseline U14 (V$_{cont}$=0.64$\pm$0.04) with a similar baseline length  and a close PA. Moreover, the visibility of the smallest baseline U34 (V$_{cont}$=0.72$\pm$0.03) is very close to the visibility of U14, implying that a relatively extended environment is seen in that particular direction ($\sim$110$^\circ$). Note that the level of the continuum visibilities from 1.95$\mu$m to 2.25$\mu$m is apparently stable within the error bars.
The visibilities decrease strongly in the Br$\gamma$ and HeI lines and the hierarchy of visibility levels is kept through the line, suggesting a similar geometry for the emitting regions, but at larger scale. 
 
An 'S-shaped' signal can be seen in the U14 and U34 differential phase curves implying that the ejecta are clearly split by the spectrograph between red and blue-shifted motions at different sky positions. The extrema of the S curves are determined by the maximum photocenter shifts of the Br$\gamma$ emission projected on the baseline directions. The radial velocities of the extrema are -\,567/661 and -\,944/850\,km.s$^{-1}$ for U34 and U14 respectively.
A strong phase jump (FWHM=500\,km.s$^{-1}$) that dominated the Br$\gamma$ phase signal of U13 and can be attributed to the crossing of a null of the visibility curve. The null signature is shifted at radial velocity -\,220$\pm$90\,km.s$^{-1}$ and seems to be mixed with a low level (amplitude $\leq$10$^\circ$, from -\,1500 to 1500km.s$^{-1}$) inverted S-shaped signal, probably responsible for the observed shift. The baseline U34 (PA=110$^\circ$) shows an inverted S-signal with extrema reaching 2500\,km.s$^{-1}$ suggesting that fast material is seen in the East-West direction. This signal is weak in the U13 and U14 baselines but is confirmed in the closure phase that exhibits a detectable asymmetry at velocities reaching -\,3000\,km.s$^{-1}$. It is worth noting that the high SNR continuum closure phases are close to zero (0.0$^\circ\pm$2.3$^\circ$), implying that the continuum source is centrally symmetric. Moreover, the integrated differential phase from U14 and U34 over the Br$\gamma$ line is very close to zero. The spectrally unresolved Br$\gamma$ emission is probably centro-symmetric, but it is difficult to claim this with confidence because of the strong phase jump of the phase from baseline U13.

\begin{table*}[]
	\centering
		\begin{tabular}{l|ccc|cc|ccc}
			\hline
			\hline
			\multicolumn{1}{c|}{Model}& \multicolumn{3}{c|}{Uniform Ellipse}&\multicolumn{2}{c|}{Gaussian Ellipse}&\multicolumn{3}{c}{Uniform Ring}\\
			\hline
\multicolumn{1}{c|}{Parameter}&\multicolumn{1}{c}{PA}&\multicolumn{2}{c|}{$\Theta$ (mas)}&\multicolumn{2}{c|}{FWHM (mas)}&\multicolumn{1}{c}{PA}&\multicolumn{2}{c}{$\Theta$ (mas)}\\
		& (degree)  &  Major& Minor & Major  & Minor&(degree) &Major & Minor \\
			\hline
Continuum 2.13$\mu$m& 142$\pm$5& 4.9$\pm$0.4& 3.0$\pm$0.3& 3.1$\pm$0.2& 1.9$\pm$0.3& 143$\pm$5 & 3.7$\pm$0.3& 1.9$\pm$0.2 \\
Br~$\gamma$~2.17$\mu$m& 140$\pm$4& 7.5$\pm$0.3& 4.7$\pm$0.4& 4.9$\pm$0.3&2.9$\pm$0.4&145$\pm$5 &5.6$\pm$0.4 & 2.9$\pm$0.3\\
He~I~2.06$\mu$m&140$\pm$8 &9.5$\pm$0.7 & 5.0$\pm$0.9& 6.3$\pm$0.6&3.6$\pm$0.8& 177$\pm$15 &6.8$\pm$0.7& 3.2$\pm$0.5\\
			\hline
		\end{tabular}
		\caption{\label{tab:geom} Results from the fit of the continuum and core of the lines by means of geometric model fits. Note that these models are centro-symmetric and have a zero closure phase as observed for the continuum, but not for the lines. The quoted errors are at the 1$\sigma$ level.}
\end{table*}

\section{Model fitting: continuum}

\subsection{Spectrally dispersed AMBER/VLTI observations}
We used simple uniform and Gaussian ellipses to fit the continuum visibilities and also the visibilities in the line center, for which the continuum contribution is at most 25\% of the line flux. Taking into account the error bars of the measurements, good quality fit parameters of the models can be found in Table 1. For each model, there are three visibilities and three free parameters: the major axis diameter $\theta_a$, the position angle and the eccentricity.
 The striking result is the consistency between the angles derived in the continuum and in lines, and their strong flattening. The ratio of the minor and major axis of the continuum and the Br$\gamma$ line ellipses are equivalent ($\sim$0.6) whereas the one for the HeI line ellipse is smaller ($\sim$0.52).

A binary model was also tested given the striking differences of visibility levels between the baselines and because Monnier et al (2006a) have interpreted data on RS\,Oph obtained with several broadband optical interferometers in terms of a binary model with moderate flux ratio (separation 3.13$\pm$0.12\,mas, position angle 36$^\circ$$\pm$10$^\circ$, brightness ratio 0.42$\pm$0.06). Their uv-coverage is very elongated such that the parameters of the model are not well constrained. The binary models in Monnier et al. were tested upon the VLTI data. Good fits to all visibility amplitudes could be found but the continuum closure phase measured by AMBER was always inconsistent by more than at least 7$\sigma$ compared to the ones determined from a broad range of parameters based on their binary model. Moreover, the allowed binary parameters are also restricted by the weak changes of the continuum visibilities from 1.95 to 2.26$\mu$m. We do not think binary models are physically plausible given the small fraction of the early outburst flux that is expected to come from (or near) the secondary (see discussion). 

The continuum model is almost certainly centro-symmetric. As a consequence, we restricted our interpretation to the use of models derived from the single object approach. An intermediate model that is also able to provide satisfactory fits is a ring model. Such a model is natural in view of the impressive radio image obtained with the VLBA array at t=13.8d showing a moderately flattened ring (O'Brien et al. 2006). However, this model is more complex involving 4 parameters. In order to keep the model simple, the width of the ring was kept fixed to 0.05\,mas. Due to the large error bars of the HeI line data, several models were possible and only the smallest extension was kept. The best parameters are also included in Tab.1. Such a model could help to explain some apparently contradictory measurements in the frame of the ellipse modes between VLTI, Keck and IOTA interferometers (see discussion).

\subsection{Broad-band IOTA and Keck observations}
The recording of a single triplet of observations prevented us getting a clear image of this spatially complex outburst, and we could try to include the data recorded by the Keck and IOTA interferometers to improve the picture of the infrared source. Between t=4 and t=11 days, several triplets of baselines were recorded with the IOTA interferometer operating in H band. The nova is almost unresolved for the smallest IOTA baselines and the information being mostly carried by the longest ones ($\sim$20m, PA ranging from 20$^\circ$ to 50$^\circ$). There is also a unique long baseline record from the Keck Interferometer at t=4\,days, with a projected baseline (B=84.1m, PA=40.3$^\circ$) very similar to our U14 one. Their $uv$ coverage is strongly biased toward the direction of the minor axis of the ellipses presented in this paper although the extension of their symmetrical Gaussian models is much larger than the minor axis of the Gaussian fits. The Keck visibility measured at t= 4\,days is 0.45$^{+0.05}_{-0.06}$, to be compared with the U14 continuum visibility of 0.64$^{+0.03}_{-0.04}$ i.e. about 40\% larger. This discrepancy is larger than the error bars of the measurements and can, at least partly, be attributed to the lack of spectral resolution of the Keck measurement, for which the high level of continuum visibilities and the lower level line signals are mixed. The Keck observations were carried out with a K-band filter centered at $\lambda_0$=2.18$\mu$m with $\Delta \lambda=0.3$$\mu$m. An estimation of this effect can be extracted from the low SNR data that cover more than half of the K band (1.93-2.26$\mu$m), and include three strong lines (Br$\gamma$, He~I~2.058 and Br$\delta$) that represent more than 25\% of the flux in this band (see Fig.3). Considering that the 2.25-2.33$\mu$m regions devoid of strong emission lines and that the Keck observations were secured slightly earlier than the VLTI observations, when the line emission was lower, we can estimate the flux emission in their band to about 20\%. The mean U14 visibility over the spectral range 2.03-2.33$\mu$m (taking into account the 2.26-2.33$\mu$m continuum level) is 0.52 (20\% lower than the continuum value), and close to the Keck value. The discrepancy, still above the upper value of the Keck estimated visibility (0.5), may be significant considering that the AMBER observations were conducted when the source was slightly more extended, 1.5 day after the Keck ones, and with a longer projected baseline, implying that AMBER should have observed a systematically lower visibility.

It must be stressed that the effect of the lines in broad-band measurements depends in a complex manner on their number, flux, on the position angle and projected length of the baseline and indeed, on the (time variable) geometry of the source. In our case, the mean U34 visibility is only 15\% lower than the continuum one while the mean U13 visibility is more than 40\% lower. The H band ($\lambda_0$=1.65$\mu$m, $\Delta \lambda=0.3$$\mu$m) IOTA observations are also probably dominated by the numerous Bracket lines merging to the Bracket jump ( ontribution of at least 15-20\% of the H band flux, Das et al. 2006a). At the time of these observations, their contribution must be accurately known and the extension of their forming regions also. The longer IOTA baselines ($\sim$20m) are oriented roughly in the direction of the minor axis of our models (20$^\circ$-50$^\circ$) but the Gaussian FWHM of 3.3mas deduced from the IOTA observations is longer than the longer axis of our continuum ellipse or longer than the minor axis of the Br$\gamma$ one. Certainly the contribution from the lines to the interferometric signal is important and the picture of the explosion in the H band seems different from the one in the K band (see for instance Evans et al. 2006, Fig.4). Without spectral resolution and with a limited $uv$-plane coverage, the understanding of the visibilities observed by optical interferometers for such a spatially complex and time-variable source is problematic and it is extremely difficult to merge the IOTA data with the others in order to get a global picture of the infrared source.

\section{Model fitting: lines}
\begin{figure}
  \begin{center}
\includegraphics[width=9cm, height=6cm]{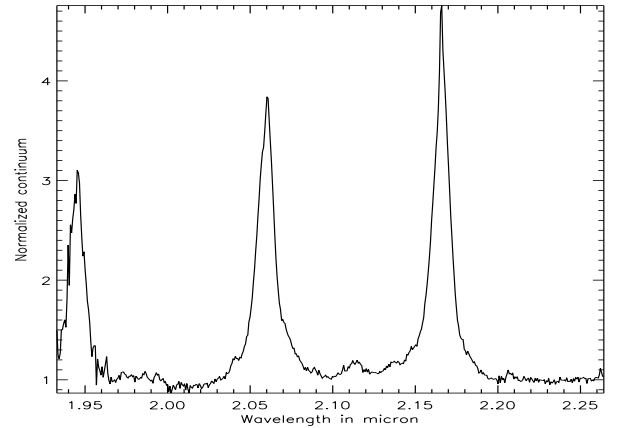}
 \end{center}
 \caption[]{K band spectrum recorded by AMBER. The lines contribute to more than 25\% of the total flux in this spectral band.
\label{fig:scales}}
\end{figure}

To exploit further the extensive spectral information and the differential phases recorded, we adopted a 'parametric imaging' approach used on the good quality Br$\gamma$ data alone, and not on the low SNR data that include the HeI line. Even though the visibilities of the low SNR data agree well with the high SNR ones, the noise level in the differential phases is such that it is difficult to estimate their true spectral shape. In each spectral channel through the Br$\gamma$ line, we first fit the visibilities with a model involving the previously defined uniform ellipse for the continuum (cf. Sect.4.1 and Tab.1) and an elliptical uniform ring for the line emission. The continuum ellipse is the reference, with differential and closure phases equal to null. Seven observables are used for the line fitting: the spectrum, three visibilities and three differential phases. The closure phase was not employed in the fitting process: it is by construction zero in the continuum model (as observed) and the line center is dominated by the phase jump of baseline U13 corresponding probably to a null crossing that can hardly be reproduced by our simple model. The flux of the rings is determined by the line flux (i.e. normalized flux minus one) of the observed spectrum. Some tests were performed showing that the position angles and eccentricity of the rings are predominantly found near the value of 140$^\circ$ and 0.6 (as for the continuum and center of the line fits reported in Tab.1). These values were kept fixed during the fitting procedure so that the continuum and rings share the same position angle for the major axis and the same ratio between major and minor axis. The main free parameters of these 'axi-symmetric' rings are their major axis and width. The fits were performed empirically without using an automatic convergence procedure, as they were providing too much channel to channel instabilities. Consequently, the proposed solution is by no means the unique and the best one. The widths of the rings are poorly constrained: their values were first chosen to slowly evolve between 0.8~mas in the wings and 1.1~mas in the line core, but other solutions can be found that were favored in the second step of the fitting process to improve the fitting of the differential phases. The range of major axis of the rings is 4.4-7.6\,mas, in between the extrema values from the continuum and Br$\gamma$ line models of Tab.1. 

In a second step, the differential phase information is taken into account by modulating the elliptical ring brightness as a function of azimuth by a sinusoid. The model is no longer axi-symmetric, and new fits are performed using the uniform ring models as a starting point. This method is inspired by the 'skewed rings' extensively introduced by Monnier et al. (2006a). The result is an empirical 2D projected kinematical map (see Fig.5) that fits all the interferometric observables and perfectly accounts for the spectrum by construction (Fig.4). The additional parameters are the amplitude of the modulation ('skew' in Monnier et al. 2006a) and the position angle of the modulation ('skew-PA'). A fit of the ring parameters is performed for each velocity interval and the resulting parameters are convolved with the spectrograph resolution (3 pixels).

\begin{figure}
  \begin{center}
\includegraphics[width=9.cm]{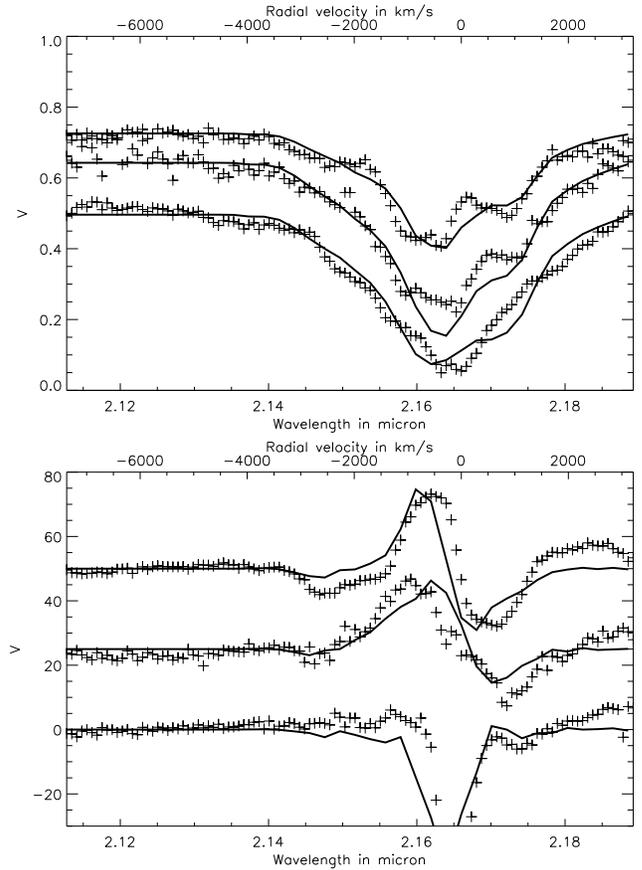}
 \end{center}
 \caption[]{Example of the fit of the visibilities (upper panel) and differential phases (lower panel) with the skewed ring models (see text and Fig.5). The fit was performed in two iterative steps. First, the visibilities are fitted using the uniform continuum and a ring for the line whose flux is proportional to the line emission at the considered spectral channel. In a second step, a sinusoidal perturbation of the ring emission (the skew) is introduced to modulate the differential phase of the model. The process is iterated several time. The smoothest solution were favored but the data are obviously more rich, evidencing the complex geometry and kinematics of the source. 
\label{fig:fit}}
\end{figure}

\begin{figure*}
  \begin{center}
\includegraphics[width=17cm]{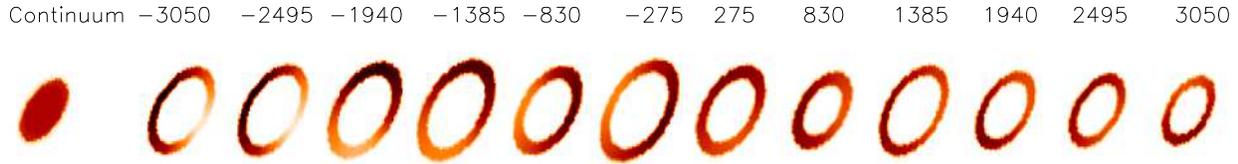}
 \end{center}
 \caption[]{Parametric model of the data involving a  uniform ellipse for the continuum (left, the size of the major axis is 4.9\,mas, other parameters can be found in Table.1) and a 'skewed ring' that account for both dispersed visibilities and differential phases (other panels, the continuum ellipse is not shown but taken into account in the calculations). The labels are expressed in km.s$^{-1}$. The images at each velocity channel are normalized and do not reflect the relative fluxes in the different velocity channels. The number of images shown is limited: there are about 80 visibility measurements in the Br$\gamma$ line and 35 independent spectral channels.
\label{fig:model}}
\end{figure*}
The skew-PA was well-constrained by the differential phases and only a small range of angles could account for the three differential phases simultaneously. The 'skew' was much less constrained due to some degeneracy with the size and thickness of the rings: this parameter controls the amplitude of the phase signal, weighted by the line contribution normalized to the continuum. In the example shown in Fig.4 and 5, the skew was limited to 0.3 from 3050km.s$^{-1}$ to -215km.s$^{-1}$, and then increases steadily to reach 0.8 for channel -\,3050km.s$^{-1}$ in order to account for the strong phase effect still seen at this velocity whereas the line flux represents less than 5\% of the continuum flux. The limits of this simple approach are rapidly reached. The 'skew' necessary to account for the data was close to 1 for these particular channels (-\,3050km.s$^{-1}$ and beyond) and the fit of the differential phase of baseline U34 was not satisfactory. At these wavelengths, the line contribution is less than 10\% of the continuum, but the interferometric signal is still large. This is an indication that the ring model is probably inadequate. The line emission, at the origin of the large photocenter shift witnessed by the baseline U34 is probably located farther than the spatial limits of the ring imposed in our model and should be replaced by a point or a jet-like feature off-centered at PA=270$^\circ$. This is also probably true for the channel 3050km.s$^{-1}$ but this is less visible in our data. Furthermore, the strong phase jump seen in the U13 baseline data is not accounted for directly with this simplistic model. Nevertheless, the model is able to reproduce the global level of the observed visibilities and differential phases. This model confirms the qualitative interpretation previously presented. Two expanding velocity fields are present, a 'slow' one between -1800 and 1800\,km.s$^{-1}$ and a 'fast' one, between -3000/-1800km.s$^{-1}$ and 1800/3000km.s$^{-1}$. The slow velocity field is modeled with rings of moderate skewness and slowly changing skew-PA that are needed to account for the difference of the velocity of maxima of the S-shaped phases (skew-PA from 55 to 110$^\circ$, modulus 180$^\circ$). The phases are probably accounted for by a model of a ring-like structure in expansion, that may correspond to the equatorial structure of the model from O'Brien et al. (2006, their Fig.4). Beyond 1800\,km.s$^{-1}$, the fast velocity field is required to explain the flip of the differential phases. Since the signal is dominantly carried by the U34 baseline, the skew-angle of the fast component is well defined in the East-West direction (skew-PA=90-270$\pm$5$^\circ$, stronger skewness). The geometry of this fast component differs significantly from the slow one and is not less convincingly accounted for within the frame of our model.

\section{Discussion}
\subsection{An expanding fireball seen by AMBER}
Thanks to the long baselines of the VLTI and the spectral resolution of AMBER, we were able to present an approximate picture of the outburst at day 5.5. An expanding fireball is seen, in the continuum and also in the lines. The lines form a kind of Str\"omgren sphere, the Br$\gamma$ line forming region is less extended than the He~I~2.06$\mu$m one, and about twice as large as the continuum forming region. The physical mechanism for the formation of the He~I~2.06$\mu$m (2$^1$S-2$^1$P) line depends on the level of ultraviolet continuum from the nova, the optical depth of the He~I~585$\AA$ (1$^1$S-2$^1$P, $h\nu \sim 21ev$) line, the decay of higher $^1$D levels, and also on the helium abundance in the ejecta (cf. Shore et al. 1996, Anupama et al. 1989, Evans et al. 1988, Whitelock et al. 1984). The level of the line flux and its spatial extension imply that the line is formed in or even beyond the fastest ejecta in an optically thick layer illuminated by the strong UV flux from the outburst. It must be noted also that the VLTI observations were performed close to the end of the phase of the outburst dominated by the ejecta, estimated at about t=5-6 days (Sokoloski et al. 2006, Bode et al. 2006, Das et al. 2006a), after which the shock velocity declines rapidly. 
At that time, the swept up mass is equivalent to the ejected shell mass.

The closure phase measured in the continuum with the 44m to 86m baselines excludes the Monnier et al. binary model for the earliest stage of the outburst, when the K band magnitude is still in the order of 4-5, i.e. 2.5-1.5 magnitude brighter than the system magnitude at quiescence (Evans et al. 1988). The closure phase value, very close to zero, implies that the continuum emission is centrally-symmetric. It must be stressed however that any model involving a binary component will be appropriate as soon as the near-IR flux is no longer dominated by the ejecta i.e. when the flux from the red giant is a non-negligible fraction of the total flux of the system. In the Hachisu \& Kato (2000, 2001) model, the irradiated face of the red giant and the accretion disk have to be taken into account from days 4 after outburst, near our time of observation, mostly to account for the UV and optical fluxes. As the ejecta expand rapidly, several effects arise: their flux drops and their correlated flux (i.e. the flux detectable by an interferometer) drops also due to the large extensions reached (18\,mas at day 25 at 1000km.s$^{-1}$ and assuming D=1.6kpc). However, at day 5.5 the K band magnitude was about 3-3.5 (Evans et al. 1988) bearing in mind that the magnitude of RS\,Oph at quiescence is 6.5 (flux that can be almost fully attributed to the red giant). It is hard to conceive that the near-IR flux was distributed such that 40\% originates in the vicinity of the red giant as in the model of Monnier et al. (2006). The alignment of the continuum and line main direction in the frame of the ellipse models is a further argument for an expanding fireball.

\subsection{Comparison with radio observations}

\begin{figure}
  \begin{center}
\includegraphics[width=7.5cm]{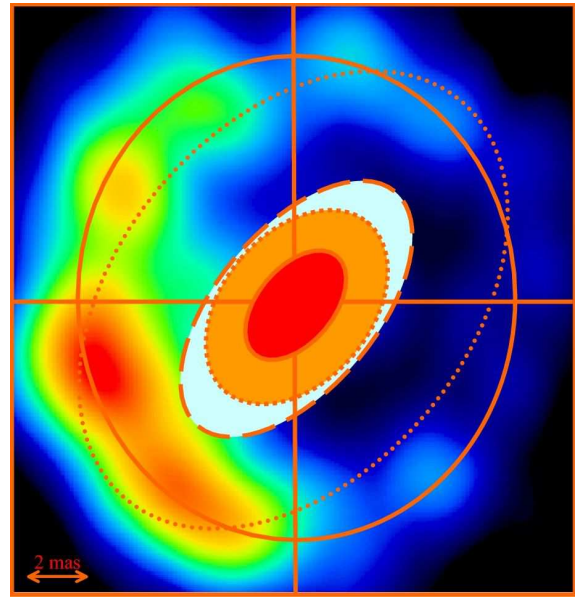}
 \end{center}
 \caption[]{Sketch of the near-IR ellipses extensions compared with the radio structure observed at t=13.8d (thick extended ring, O'Brien et al. 2006). The continuum ellipse is delimited by the solid line, the ellipse that corresponds to the core of Br$\gamma$ by the dotted line and the one corresponding to the core of HeI by the dashed line. The small dotted line delimit the Br$\gamma$ ellipse scaled at t=13.8d. North is up, East left.
\label{fig:scales}}
\end{figure}

A good complement to this optical interferometry study is the spatially resolved radio-interferometry map obtained at t=13.8 day (O'Brien et al. 2006), shown in Fig.6 together with the best uniform ellipse models. In the ellipse interpretation, the K band continuum and line emission are highly flattened and oriented in the same direction, at $\sim$135$^\circ$. This particular direction has never been reported in the literature (mostly radio observations) and is in contrast with the radio image, dominated by an almost circular structure extended slightly N-S, significantly brighter on its eastern side. \footnote{The ratio between the E-W minor and N-S major axis suggests an inclination of $\sim$35-40$^\circ$}. Their interpretation is that this is the waist of a bipolar structure extended E-W and tilted at an angle 20-40 degrees to the line of sight. The angular diameter of the radio structure at t=13.8d is 17\,mas, scaling down to 6.8\,mas at t=5.5\,d. This value is close to the Br$\gamma$ line extension, and is smaller than the HeI one. On the one hand, the expansion velocity $|v_{exp}|$ of the radio ring was determined precisely to 1730~km.s$^{-1}$ from North-South flux slices (O'Brien et al. 2006, Fig.2). This value is within the range of radial velocities of our 'slow' velocity field $|v_{rad}|$\footnote{Note that if the velocity field was purely in the equatorial plane, then the maximum radial velocity observable should be around 1400km.s$^1$. This is in agreement with the radial position of the peaks of the S-shape differential phase ranging from 600 to 1500km.$^1$ tracing the bulk of the emission from this structure (cf. Sect.3).}. One the other hand, the radio flux seen at t=13.8 day is a simple structure located in the equatorial plane of the system: the fact that we see the 'fast' velocity field in the Br$\gamma$ line implies that the interferometer sees a more complex structure projected on the sky not limited by the equatorial emission. This is probably also true for the continuum emission (Evans et al. 1988, Evans et al. 2006). In the frame of the ejection model from O'Brien et al. this means that both the equatorial waist and the bipolar lobes may contribute to the shape inferred from AMBER data result. As a consequence, it is not surprising that the global shape and position angle of the near-IR source seen by AMBER differ from the radio one.

The radio ring is rather clumpy and dominated by a few bright structures. In particular, the direction of the brightest radio clumps (PA$\sim$110-150$^\circ$) coincides with the direction of our correspondingly more flattened K band continuum, Br$\gamma$ and HeI structures. The origin of such an asymmetry could tentatively be questioned as the effect of particular configuration of the WD and the RG at the moment of outburst.
However, based on the ephemeris of Fekel et al. and assuming a line of nodes at PA$\sim$177$^\circ$ (Taylor et al. 1989, O'Brien et al. 2006), the position of the red giant should be close to PA$\sim$150-170 at phase 0.97.
It is obvious that at least a strong perturbation of the flow at t=5.5d is encountered near the red giant photosphere that has potential effects, especially in the near-IR. The WD-RG separation is typically 1\,mas (assuming D=1.6kpc), and the red giant diameter 0.4 mas: seen from the ejection center, the covered angle is about 15-30$^\circ$. At t=5.5\,d, the K band emitting ejecta spanned about 2 times the WD-RG distance and probably kept a strong signature of the perturbation by the RG, which should eventually dilute as the ejecta expand. This signature may also be imprinted in the earliest high resolution radio images. Finally, the E-W direction of the fast Br$\gamma$ emitting regions seen with our differential phase coincides with the direction of the jet-like structure developing in radio images at day 21.5 and after, but the near-IR radial velocities are about two times slower than the apparent motion of the structure in the radio.

\section{Conclusion}
We have reported the early near-IR AMBER interferometric observation of the outburst of RS\,Oph, a spectrally impressive but spatially limited dataset from the point of view of image reconstruction. These observations performed 5.5\,days after the outburst provided an estimation of the extent of the continuum, Br$\gamma$ and He~I~2.06$\mu$m forming regions and some physical constraints on the ejection process as seen in the near-IR. The global picture that emerges in view of the consistency between the shape of the continuum and line forming regions, despite various physical process at their origin, is a non-spherical fireball at high-velocity expansion. Our results represent a good complement of the extensive radio, infrared and X-ray observations.

In order to study carefully fast evolving events like nova, supernovae, and other kinds of outbursting sources, we have shown that it is critical that spectral and imaging capabilities must be available simultaneously. The $uv$ coverage has to be such that a good picture of the ejection must be recorded in one-two nights at most and a minimum spectral resolution is mandatory to, at least, isolate the line signal from the continuum one. This difficult task can potentially be achieved with the current capabilities of the AMBER instrument, but it fits more easily the goals of the second generation NIR and MIR VLTI instruments under study, able to recombine light from at least 4 telescopes.

The near-IR picture of RS Oph evolves in a complex way: it is a mix of increasingly extended ejecta emitting a continuum and also many lines whose fluxes decrease with time, and a very compact source (at the assumed distance of 1.6kpc) that becomes dominant during quiescence. In order to develop models of the nova outburst, it is of prime importance to constrain further the fundamental parameters of the system, in particular its distance, inclination and parameters of the orbit in the plane of the sky. This can only be achieved by optical interferometers and we advocate a monitoring of the system by the VLTI in the course of the orbital period of 455 days, that should provide an accurate determination of the component masses. This implies to use the longest VLTI baselines available with the UTs (100-130m) in the J and H band preferentially, with the low resolution mode (R$\sim$30). With this range of baseline, the individual components of the system can be considered as unresolved by the interferometer. Such a study is possible now that the red giant signature is visible again through the CO signature in the K band as observed by Das et al. 2006b at t=243 days.


\begin{acknowledgements}
These observations were made at ESO through a Director's Discretionary Time procedure. We are grateful to those who rapidly took the decision of acceptance allowing us to record these impressive data and we also thank Christian Pollas for having provided the very first information on this event.
We thank John Monnier and Richard Barry for having provided the IOTA and Keck data. M.F. B. is grateful to the UK PPARC for provision of a Senior Fellowship. We are grateful to the referee, Steven Shore for comments that improved the manuscript.

\end{acknowledgements}


\begin{thebibliography}{}

\bibitem{} Anupama, G.C. \& Prabhu, T.P., J. Astrophys. \& Astron., 10, 237

\bibitem{} Bode, M.F., O'Brien T.J., Osborne, J.P. et al., 2006, ApJ, 652, 629

\bibitem{} Bode, M.F., ed. {\it RS Ophiuchi (1985) and the recurrent nova phenomenon}, VNU Science Press,
Utrecht, 1987

\bibitem{} Das, R.K., Banerjee, D.P.K. and Ashok, N.M., 2006a, astro.ph 0611254

\bibitem{} Das, R.K., Ashok, N.M. and Banerjee, D.P.K., 2006b, CBET 730

\bibitem{} Evans, A., Callus, C.M., Albinson, J.S. et al. 1988, MNRAS, 234, 755

\bibitem{} Evans, A., Kerr, T., Yang, B. et al., 2006, astro.ph, 0609394

\bibitem{} Fekel, F.C., Joyce, R.R., Hinkle, K.H. and Skrutskie, M.F., 2000, AJ, 119, 1375

\bibitem{} Gehrz, R.D., Truran, J.W., Williams, R.E. and Starrfield, S., 1998, PASP, 110, 3

\bibitem{} Gehrz, R.D. 1988, ARA\&A, 26, 377

\bibitem{} Hachisu, I. \& Kato, M. 2000, ApJ, 536, L93

\bibitem{} Hachisu, I. \& Kato, M. 2001, ApJ, 558, 323

\bibitem{} Hachisu, I. \& Kato, M. 2006, ApJ, 642, L53

\bibitem{} Hachisu, I., Kato, M., Kiyota, S. et al. 2006, ApJ, 651, L141

\bibitem{} Hirosawa, K., 2006, IAUC 8671

\bibitem{} Hjellming, R.M., van Gorkom, J.H., Seaquist, E.R. et al. 1986, ApJ, 305, L71


\bibitem{} Lane, B.F., Retter, A., Eisner, J.A. et al., 2005, AAS, 20710408

\bibitem{} Lane, B.F., Retter, A., Thompson, R.R. and Eisner, J.A. 2005, ApJ, 622, L137
	
\bibitem{} Leinert, Ch., Graser, U., Przygodda, F. et al., 2003, Ap\&SS, 286, 73

\bibitem{} Lloyd, H.M., Bode, M.F., O'Brien, T.J. and Kahn, F.D., 1997, MNRAS, 284, 137

\bibitem{} Monnier, J.D., Berger, J.-P., Millan-Gabet, R., 2006b, ApJ, 647, 444

\bibitem{} Monnier J.D., Barry, R.K., Traub, W.A. et al., 2006a, ApJ Letter, 647, L127

\bibitem{} O'Brien, T.J., Bode, M.F., Porcas, R.W. et al., 2006, Nature, 442, 279

\bibitem{} Padin, S., Davis, R.J., Bode, M.F. 1985, Nature, 315, 306

\bibitem{} Petrov, R.G. and The AMBER Consortium, 2003, Ap\&SS, 286, 57

\bibitem{} Quirrenbach, A., Elias, N.M., Mozurkewich, D., et al. 1993, AJ, 106, 1118

\bibitem{} Richichi, A. \& Percheron, I. 2005, A\&A, 434, 1201

\bibitem{} Richichi, A., Percheron, I. and Khristoforova, M., 2005, A\&A, 431, 773

\bibitem{} Shore, S. Kenyon, S.J., Starrfield, S. and Sonneborn, G. 1996, ApJ, 456, 717

\bibitem{} Snijders, M.A.J. 1987, in {\it RS Ophiuchi (1985) and the Recurrent Nova Phenomenon, ed. M.F. Bode}, VNU Science Press, Utrecht,51

\bibitem{} Sokoloski, J.L., Luna, G.J.M., Mukai, K. et al., 2006, Nature, 442, 276

\bibitem{} Starrfield, S., Truran, J. W. and Sparks, W.M., 2000, New Astronomy Reviews, 44, 81

\bibitem{} Starrfield, S., Timmes, F.X., Hix, W.R. et al., ApJ, 612, L56 

\bibitem{} Tatulli, E., Millour, F., Chelli, A., et al. 2006, A\&A, accepted

\bibitem{} Taylor, A.R., Davis, R.J., Porcas, R.W. and Bode, M.F. 1989, MNRAS, 237, 81

\bibitem{} Whitelock, P.A., Carter, B.S., Feast, M.W. et al., 1984, MNRAS, 211, 421

\bibitem{} Wood-Vasey, W.M. and Sokoloski, J.L., ApJ, 645, L53



\end{thebibliography}
\end{document}